%% file: main.tex
\documentclass[12pt]{article}

\usepackage[utf8]{inputenc}
\usepackage{amsmath, amsthm, amssymb, amsfonts}
\usepackage{import}
\usepackage[authoryear, comma]{natbib}

\usepackage{fullpage}
\usepackage{authblk}
\usepackage[bibliography=common]{apxproof}
\usepackage{import}
\usepackage{mathtools}
\usepackage{tikz}
\usepackage{float}
\usepackage{enumitem}
\usepackage{bbm}
\usepackage[colorlinks=false,pagebackref=true, hidelinks]{hyperref}
\usetikzlibrary{calc,patterns,angles,shapes, positioning, intersections, quotes}

\newtheoremrep{thm}{Theorem}
\newtheoremrep{lem}{Lemma}
\newtheoremrep{prop}{Proposition}

\newtheoremrep{proposition}{Proposition}

\usepackage{graphicx}
\newtheorem{corollary}{Corollary}

\theoremstyle{definition}

\newcommand{\R}{\mathbb{R}}

\allowdisplaybreaks

\date{\today\\\small{\href{https://goelsumit.com/files/auctions_efficiency.pdf}{(Link to latest version)}}}

\begin{document}

\title{An efficiency ordering of k-price auctions under complete information
}

\author{Sumit Goel\thanks{NYU Abu Dhabi; \href{mailto:sumitgoel58@gmail.com}{sumitgoel58@gmail.com}; 0000-0003-3266-9035}
\quad Jeffrey Zeidel\thanks{NYU Abu Dhabi; \href{mailto:jrz8904@nyu.edu}{jrz8904@nyu.edu}}}

\maketitle        

\import{files/}{abstract.tex} 
\import{files/}{intro.tex} 
\import{files/}{model.tex} 
\import{files/}{results.tex}

\newpage
\bibliographystyle{ecta}

\bibliography{refs}

\end{document}

%% file: files/abstract.tex
\begin{abstract}
We study $k$-price auctions in a complete information environment and characterize all pure-strategy Nash equilibrium outcomes. In a setting with $n$ agents having ordered valuations, we show that any agent, except those with the lowest $k-2$ valuations, can win in equilibrium. As a consequence, worst-case welfare increases monotonically as we go from  $k=2$ (second-price auction) to $k=n$ (lowest-price auction), with the first-price auction achieving the highest worst-case welfare. 
\end{abstract}

%% file: files/intro.tex
\section{Introduction}

We study $k$-price sealed-bid auctions in a complete information environment with $n$ agents who have strictly ordered valuations. In a $k$-price auction, all $n$ agents submit their bids,  the highest bidder wins the object (with ties broken in favor of the agent with the highest valuation), and pays the $k$th highest bid. We fully characterize the set of pure-strategy Nash equilibrium outcomes for every $k$-price auction. \\

For $k \in \{2, \dots, n\}$, we show that any of the top $n-(k-2)$ valuation agents can win in equilibrium, while the bottom $(k-2)$ agents can never win. In other words, the second-price auction can be won by any of the $n$ agents, the third-price auction by any of the top $n-1$ agents, and so on, until we reach the lowest-price auction ($k=n$), which can only be won by the top two agents. We further show that in the first-price auction ($k=n+1$), only the top agent can win. This equilibrium characterization reveals a natural ordering of $k$-price auctions in terms of their worst-case allocative efficiency: the worst-case equilibrium allocation becomes strictly more efficient as $k$ increases from $2$ to $n+1$. \\

In closely related work, \citet*{tauman2002note} and  \citet*{mathews2017note} also study $k$-price auctions in complete information environments. Under the restriction to pure strategies that are not weakly dominated, \citet*{tauman2002note} shows that for any $k$-price auction, only the top agent can win in equilibrium. Subsequently, \citet*{mathews2017note} construct an equilibrium in mixed-strategies where the top agent does not win. In comparison, we characterize all pure-strategy Nash equilibrium outcomes for all $k$-price auctions and obtain an ordering of these auctions based on their worst-case allocative efficiency.\footnote{Other work on $k$-price auctions has focused on incomplete information settings (\citet*{kagel1993independent,monderer2000k, monderer2004k, mezzetti2009auctions, azrieli2012dominance, mihelich2020analytical, skitmore2014km}).}

%% file: files/model.tex
\section{Model}

A seller is selling an indivisible object to a set $N=\{1, \dots, n\}$ of agents. Each agent $i\in N$ has a valuation $v_i>0$ for the object, and we assume that $$v_1>v_2>\dots>v_n.$$
We further define $v_{n+1}=0$. The valuations are assumed to be common knowledge. \\

The object is sold using a sealed-bid $k$-price auction. Each agent $i\in N$ simultaneously submits a non-negative bid $b_i\in \R_+$. The object is awarded to the agent who submits the highest bid, with ties resolved in favor of the agent with the highest valuation.\footnote{Previous work has typically assumed that ties are broken uniformly at random (\citet*{tauman2002note, mathews2017note}). We note that our results for $k$-price auctions with $k\in \{2, \dots, n\}$ continue to hold under either tie-breaking rule. The choice of tie-breaking rule affects only the first-price auction and makes no substantive difference to the main conclusions. Our assumption allows us to discuss the first-price auction alongside the other $k$-price auctions in a unified way, aids exposition more generally, and is natural given our focus on welfare. Additionally, we note that from the seller’s perspective, implementing this tie-breaking rule requires only knowledge of the ordinal ranking of agents’ valuations, not their exact values.}
The winner pays the $k$th highest bid, denoted $b_{(k)}$, and all other agents pay zero. The utility of agent $i\in N$ at bid profile $b=(b_1, \dots, b_n)$ is given by:
$$
u_i(b)= \begin{cases}v_i-b_{(k)} & \text { if } i=\min \{j \in N: b_j=b_{(1)}\}, \\ 0 & \text { otherwise. }\end{cases}
$$
We characterize pure-strategy Nash equilibrium outcomes of the $k$-price auction for all $k \in \{1, 2, \dots, n\}$, where $k=1$ denotes the first-price auction, $k=2$ the second-price auction, and so on, with $k=n$ representing the lowest-price auction.\footnote{Since our analysis focuses on pure-strategy equilibria, our characterization results actually depend only on the ordinal ranking over bid profiles induced by the representation $u_i$. Specifically, we can define the outcome set for each agent as $X=\mathbb{R}_{+} \cup\{-1\}$, where an outcome $x \in \mathbb{R}_{+}$ represents the payment made by the agent when it wins the object, and $x=-1$ denotes not winning. An agent $i \in N$ with valuation $v_i>0$ is then represented by a preference relation $\succ_i$ over $X$, such that $x \succ_i y$ for all $x<y \in \mathbb{R}_{+}$, and $-1 \sim_i v_i$. Our results hold as long as the utility representation $u_i$ is consistent with $\succ_i$ for all $i\in N$.} For notational convenience, we sometimes also use $k=n+1$ to refer to the first-price auction. 


%% file: files/results.tex
\section{Results}

We first characterize equilibrium outcomes from the seller's perspective. Specifically, we show that under the $k$-price auction, the seller's revenue must lie within the interval $[v_{n-(k-3)}, v_1]$, and any revenue within this interval is possible. 
\begin{proposition}
\label{prop:seller}
Consider a $k$-price auction with $k\in \{2, \dots, n+1\}$. There exists a pure-strategy Nash equilibrium in which the seller's revenue is $p$ if and only if $$p\in [v_{n-(k-3)}, v_1].$$ 
\end{proposition}
\begin{proof}
Fix any $k$-price auction. We first show that in equilibrium, the revenue $p\in  [v_{n-(k-3)}, v_1]$. Suppose towards a contradiction that $b$ is an equilibrium profile and $p = b_{(k)}\notin [v_{n-(k-3)}, v_1]$. 
Notice first that if $p>v_{1}$, the winner's utility is negative, which is not possible in equilibrium (since the winner can deviate by bidding $0$). Thus, $p<v_{n-(k-3)}$. We analyze why this is not possible separately for different $k$. 
 
    \begin{enumerate}
        \item $k=2$: For the second-price auction, $p<v_{n-(k-3)}$ simplifies to  $p<v_{n+1}=0$. But $p=b_{(2)}\geq 0$, which is a contradiction.
        \item $k\in \{3, \dots, n\}$: For this $k$-price auction, $p<v_{n-(k-3)}$ and $p=b_{(k)}$ imply that
    \begin{enumerate}
        \item At least $(n-(k-3))$ agents have valuation $>p$ (as $v_1>\dots>v_{n-(k-3)}>p$).
        \item At least $k$ agents bid $\geq p$ at profile $b$.
    \end{enumerate}
    It follows then that there are at least three distinct agents with valuations strictly greater than $p$, each bidding at least $p$. At least two of these three agents receive utility $0$. At least one of these two agents can deviate by bidding $> b_{(1)}$, and receive strictly positive utility. This contradicts $b$ being an equilibrium.
        \item $k=n+1$: For the first-price auction, $p<v_{n-(k-3)}$ simplifies to $p<v_2$. Notice that at least one of agents $1$ and $2$ receive utility $0$, and this agent can deviate by bidding in $(b_{(1)}, v_2)$, and receive strictly positive utility. This contradicts $b$ being an equilibrium. 
        
    \end{enumerate}

It follows that in any equilibrium, $p\in [v_{n-(k-3)}, v_1]$. \\

Now we show that for any $p\in [v_{n-(k-3)}, v_1]$, there exists an equilibrium in which the revenue is $p$. We construct such an equilibrium separately for different $k$. 

\begin{enumerate}
    \item $k\in \{2, \dots, n\}$: For this $k$-price auction, and $p\in [v_{n-(k-3)}, v_1]$, consider the bid profile
$$b=(v_1, \underbrace{p, \dots, p}_{n-(k-1)\text { agents}}, \underbrace{v_1, \dots, v_1}_{(k-2)\text { agents}}).$$
At profile $b$, agent $1$ wins the $k$-price auction (as ties are broken in favor of agent with highest valuation), and pays a price $b_{(k)}=p$ for the good. Agent $1$'s utility is $v_1-p \geq 0$ and for $j\neq 1$, agent $j$'s utility is $0$. We now verify that $b$ is indeed a Nash equilibrium. Consider agent $j\in N$. 
\begin{enumerate}
    \item $j=1$: If $b_1'\geq v_1$, agent $1$'s utility does not change. If $b_1'<v_1$, agent $1$'s utility is either $0$ or does not change (possible when $k=2$). 
    \item $j\in \{2, \dots, n-(k-2)\}$: If $b_j'>v_1$, agent $j$'s utility will be $v_j-v_1<0$. If $b_j'\leq v_1$, agent $j$'s utility remains $0$. 
    \item $j\in \{n-(k-3), \dots, n\}$: If $b_j'>v_1$, agent $j$'s utility will be $v_j - p \leq 0$. If $b_j'\leq v_1$, agent $j$'s utility remains $0$. 
\end{enumerate}
Thus, no agent has a profitable deviation, and $b$ is an equilibrium in the $k$-price auction.
\item $k=n+1$: For the first-price auction, and $p\in [v_2, v_1]$, consider the bid profile
$$b=(p, \dots, p).$$
At the profile $b$, agent $1$ wins the first-price auction (as ties are broken in favor of agent with highest valuation), and pays a price $b_{(1)}=p$ for the good. Agent $1$'s utility is $v_1-p \geq 0$ and for $j\neq 1$, agent $j$'s utility is $0$. It is straightforward to verify that $b$ is a Nash equilibrium in the first-price auction.

\end{enumerate}
Thus, for any $p\in [v_{n-(k-3)}, v_1]$, there exists an equilibrium in which the revenue is $p$.\footnote{There are many other equilibrium bid profiles that induce the same outcome. In particular, for $k\in \{2, \dots, n\}$, we can modify our $b$ so that $b_1>v_1$, and the resulting profile remains a Nash equilibrium with the same outcome, irrespective of the tie-breaking rule.
}
\end{proof}

We now characterize equilibrium outcomes from the buyers perspective. Under the $k$-price auction, we show that any agent whose valuation exceeds $v_{n-(k-3)}$ can win at any price between $v_{n-(k-3)}$ and their own valuation, and that no other outcomes are possible.

\begin{proposition}
\label{prop:buyers}
Consider a $k$-price auction with $k\in \{2, \dots, n+1\}$. There exists a pure-strategy Nash equilibrium in which agent $i\in N$ wins and pays $p$ if and only if $$v_i> v_{n-(k-3)} \text{ and } p\in [v_{n-(k-3)}, v_i].$$ 
\end{proposition}

\begin{proof}
Fix any $k$-price auction. We first show that in equilibrium, if agent $i$ wins and pays $p$, it must be that $v_i>v_{n-(k-3)}$ and $p\in [v_{n-(k-3)}, v_i]$.
Suppose towards a contradiction that $b$ is an equilibrium profile in which agent $i\in N$ with $v_i\leq v_{n-(k-3)}$ wins. From Proposition \ref{prop:seller}, it must pay $p\geq v_{n-(k-3)}$. But since the winner's utility must be non-negative in equilibrium (as it can deviate by bidding 0 otherwise), the only possibility is that $v_i=p=v_{n-(k-3)}$. 
We analyze why this is not possible separately for different $k$.
 
    \begin{enumerate}
    \item $k=2$: For the second-price auction, $v_i=p=v_{n-(k-3)}$ simplifies to $v_i=p=v_{n+1}=0$. But $v_i>0$ for all $i\in N$, which is a contradiction.
        \item $k\in \{3, \dots, n\}$: For this $k$-price auction, $v_i=p=v_{n-(k-3)}$ and $p=b_{(k)}$ imply that
    \begin{enumerate}
        \item At least $(n-(k-2))$ agents have valuation $>p$ (as $v_1>\dots>v_{n-(k-2)}>p$).
        \item At least $k$ agents bid $\geq p$ at profile $b$.
    \end{enumerate}
    It follows then that there are at least two distinct agents with valuations strictly greater than $p$, each bidding at least $p$. Further, both these agents receive utility $0$, and at least one of these two agents can deviate by bidding $> b_{(1)}$, and receive strictly positive utility. This contradicts $b$ being an equilibrium. 
    \item $k=n+1$: For the first-price auction, $v_i=p=v_{n-(k-3)}$ simplifies to $v_i=p=v_2$. But then, agent $1$ receives a utility of $0$, and it can deviate by bidding in the interval $(v_2, v_1)$, and receive strictly positive utility. This contradicts $b$ being an equilibrium.
\end{enumerate}
Thus, it must be that $v_i>v_{n-(k-3)}$. Further, if agent $i$ wins and pays $p$, it must be that $p\leq v_i$, and from Proposition \ref{prop:seller}, $p\geq v_{n-(k-3)}$. Thus, in any equilibrium where $i$ wins and pays $p$, it must be that $v_i>v_{n-(k-3)}$ and $p\in [v_{n-(k-3)}, v_i]$.\\

Now we show that for any $i\in N$ such that $v_i> v_{n-(k-3)}$ and $p\in [v_{n-(k-3)}, v_i]$, there exists an equilibrium where $i$ wins and pays $p$. We construct one separately for different $k$. 

\begin{enumerate}

\item $k\in \{2, \dots, n\}$: For this $k$-price auction, $i\in N$ such that $v_i> v_{n-(k-3)}$ and $p\in [v_{n-(k-3)}, v_i]$, consider the bid profile $b$ where 
$$b_i = v_1 \text{ and } b_{-i}=(\underbrace{p, \dots, p}_{n-(k-1)\text { agents}}, \underbrace{v_1, \dots, v_1}_{(k-2)\text { agents}}).$$
At profile $b$, agent $i$ wins the $k$-price auction (as ties are broken in favor of agent with highest valuation), and pays a price $b_{(k)}=p$ for the good. Agent $i$'s utility is $v_i-p \geq 0$ and for $j\neq i$, agent $j$'s utility is $0$. We now verify that $b$ is indeed a Nash equilibrium. Consider agent $j\in N$.
\begin{enumerate}
    \item $j=i$: If $b_i'\geq v_1$, agent $i$'s utility does not change. If $b_i'<v_1$, agent $i$'s utility is either $0$ or does not change (possible when $k=2$). 
    \item $j\in \{1, \dots, n-(k-2)\} \setminus \{i\}$: If $b_j'>v_1$, agent $j$'s utility will be $v_j-v_1\leq 0$. If $b_j'\leq v_1$, agent $j$'s utility will be either $0$ or $<0$. 
    \item $j\in \{n-(k-3), \dots, n\}$: If $b_j'>v_1$, agent $j$'s utility will be $v_j - p \leq 0$. If $b_j'\leq v_1$, agent $j$'s utility remains $0$. 
\end{enumerate}
Thus, no agent has a profitable deviation, and $b$ is an equilibrium in the $k$-price auction.

\item $k=n+1$: For the first-price auction, $i=1$ and $p\in [v_2, v_1]$, consider the bid profile $$b=(p, \dots, p).$$
It is straightforward to verify that $b$ is a Nash equilibrium in the first-price auction. 
\end{enumerate}
Thus, for any $i\in N$ such that $v_i> v_{n-(k-3)}$ and $p\in [v_{n-(k-3)}, v_i]$, there exists a Nash equilibrium where agent $i$ wins and pays $p$.
\end{proof}

Proposition \ref{prop:buyers} yields a natural ranking of $k$-price auctions in terms of their worst-case efficiency. Formally, we let
$$\underline{W}^k = \min\{v_i: \exists b \text{ such that } b \text{ is an equilibrium of the } k\text{-price auction where agent } i \text{ wins} \},$$
denote the worst-case equilibrium welfare under the $k$-price auction. 

\begin{corollary}
For $k\in \{2, \dots, n+1\}$, the worst-case equilibrium welfare of  $k$-price auction is $$\underline{W}^k = v_{n-(k-2)}.$$
Hence, 
$$\underline{W}^2<\underline{W}^3<\dots<\underline{W}^n<\underline{W}^{n+1}.$$

\end{corollary}
Thus, the second-price auction is the worst, as even the lowest-valuation agent can win, while higher values of $k$ progressively exclude lower-valuation agents from winning in equilibrium, culminating in the first-price auction, where only the highest-valuation agent can win.\footnote{From Proposition \ref{prop:seller}, the same ordering also emerges when ranking $k$-price auctions by worst-case revenue.}\\

\begin{figure}[h]
\caption{k-Price Auctions With 5 Agents}
\begin{center}
    \includegraphics[scale=.39]{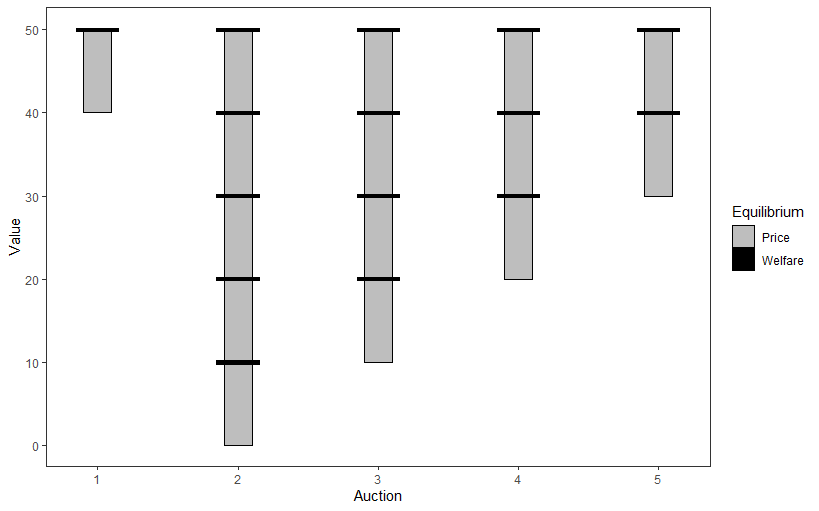} 
\end{center}
\label{fig:example}
\end{figure}

Lastly, we illustrate our results through an example with $n=5$ agents whose valuations are $v_1=50,v_2=40,v_3=30,v_4=20$ and $v_5=10$. In figure \ref{fig:example} the black bars show the possible equilibrium welfare, while the grey vertical bars show the interval of possible equilibrium prices. In the first-price auction, only agent $1$ can win at a price in $[40, 50]$. In the second-price auction, any agent $i$ can win at a price in $[0,v_i]$. And so on, until we reach the lowest-price auction, where either agent $2$ can win at a price in $[30, 40]$ or agent $1$ can win at a price in $[30, 50]$. For example, under the lowest-price auction, $b=(35, 70, 65, 60, 55)$ is an equilibrium profile in which agent $2$ wins (leading to welfare of $40$) at price $p=35$. Since the second lowest bid is above $v_1$, agent 1 is deterred from raising his bid, and since the price is above $v_3$, agents 3 through 5 are deterred from raising theirs.   

